\documentclass[sigconf,natbib=false]{acmart}

\copyrightyear{2026}
\acmYear{2026}
\setcopyright{cc}
\setcctype{by}
\acmConference[ICPC '26]{34th IEEE/ACM International Conference on Program Comprehension}{April 12--13, 2026}{Rio de Janeiro, Brazil}
\acmBooktitle{34th IEEE/ACM International Conference on Program Comprehension (ICPC '26), April 12--13, 2026, Rio de Janeiro, Brazil}
\acmPrice{}
\acmDOI{10.1145/3794763.3794800}
\acmISBN{979-8-4007-2482-4/2026/04}

\usepackage[cleanlook,british]{isodate}
\usepackage[nodraftspread,todo,coloredits]{mycommon}

\usepackage{csquotes}
\usepackage{booktabs,tabularx,multirow,multicol}
\usepackage[inline]{enumitem}
\usepackage{subcaption}
\usepackage{apastat}

\makecoloredit{\sam}{phthaloblue}
\makecoloredit{\robert}{red}
\makecoloredit{\bonita}{yellow}

\usepackage{siunitx}
\DeclareSIUnit\px{px}

\usepackage{graphicx}

\usepackage{listings}

\lstdefinelanguage{TypeScript}{
  keywords={abstract, any, as, boolean, break, case, catch, class, console, 
    const, continue, debugger, declare, default, delete, do, else, enum, export, 
    extends, false, finally, for, from, function, get, if, implements, import, in, 
    infer, instanceof, interface, keyof, let, module, namespace, never, new, null, 
    number, object, package, private, protected, public, readonly, require, return, 
    set, static, string, super, switch, symbol, this, throw, true, try, type, typeof, 
    undefined, unique, unknown, var, void, while, with, yield},
  morecomment=[l]{//},
  morecomment=[s]{/*}{*/},
  morestring=[b]',
  morestring=[b]",
  ndkeywords={class, export, boolean, throw, implements, import, this},
  sensitive=true,
}

\newcommand{\annotbox}[1]{\fcolorbox{seaborncolorblind1!40}{seaborncolorblind1!5}{#1}}
\newcommand{\aoibox}[1]{\fcolorbox{seaborncolorblind2!40}{seaborncolorblind2!5}{#1}}

\usepackage[style=acmnumeric,mincitenames=1,maxcitenames=2,sorting=nty]{biblatex}
\usepackage[nameinlink,capitalise]{cleveref}

\usepackage{tikz}
\usetikzlibrary{arrows.meta,shapes.geometric,shapes.misc,decorations,positioning}

\usepackage{xspace}
\newcommand{\tsc}{TypeScript\xspace}

\begin{document}

\title{Do Developers Read Type Information? An Eye-Tracking Study on TypeScript}

\author{Samuel W. Flint}
\authornote{Most of this work was completed while this author was a student at the University of Nebraska--Lincoln.}
\email{Samuel.Flint@dsu.edu}
\orcid{0000-0002-8023-9710}
\affiliation{
  \institution{Dakota State University}
  \department{Beacom College of Computer \& Cyber Sciences}
  \city{Madison}
  \state{SD}
  \country{USA}
}

\author{Robert Dyer}
\email{rdyer@unl.edu}
\orcid{0000-0001-9571-5567}
\affiliation{
  \institution{University of Nebraska-Lincoln}
  \department{School of Computing}
  \city{Lincoln}
  \state{NE}
  \country{USA}
}

\author{Bonita Sharif}
\email{bsharif@unl.edu}
\orcid{0000-0002-5178-7160}
\affiliation{%
  \institution{University of Nebraska---Lincoln}
  \department{School of Computing}
  \city{Lincoln}
  \state{NE}
  \country{USA}
}

\begin{abstract}
Statically-annotated types have been shown to aid developers in a number of programming tasks, and this benefit holds true even when static type checking is not used.
It is hypothesized that this is because developers use type annotations as in-code documentation.
In this study, we aim to provide evidence that developers use type annotations as in-code documentation.
Understanding this hypothesized use will help to understand how, and in what contexts, developers use type information; additionally, it may help to design better development tools and inform educational decisions.
To provide this evidence, we conduct an eye tracking study with 26 undergraduate students to determine if they read type annotations during code comprehension and bug localization in the \tsc language. 
We found that developers do not look directly at lines containing type annotations or type declarations more often when they are present, in either code summarization or bug localization tasks.
The results have implications for tool builders to improve the availability of type information, the development community to build good standards for use of type annotations, and education to enforce deliberate teaching of reading patterns.
\end{abstract}

\maketitle

\section{Introduction}
\label{sec:introduction}

Optional type annotations are an important feature of a number of programming languages, enforced to various extents.
However, they are also theorized to act as in-code documentation~\cite{lubin21:_how}, and may help developer productivity in a number of ways (at least, if they are correct)~\cite{spiza14:_type_apis}.
This is especially important, as one of the ways type annotations can help is by reducing the chance of certain kinds of errors~\cite{mayer12:_empir_study_influen_static_type,fischer15:_api_types_javas_ms_visual_studio}.
However, there is limited work that explains \emph{how} type annotations perform their theorized documentary function.
Namely, when reading code, how often do developers read type annotations, and for how long?

\tsc is gaining popularity as a statically typed language and is considered to be one of the top five commonly used programming languages globally according to the 2024 State of the Octoverse survey~\cite{octoverse2024}. We chose \tsc in this study because it is a popular language and because it allows us flexibility with placing type annotations. The goal of this study is to determine if developers read and actively reference type annotations in \tsc, while they are working on comprehension and bug localization tasks. We define reference behavior as reading lines containing declarations or annotations, or referring back to them (re-reading). 
Eye tracking allows us to specifically track developer reading behavior, which is strongly implicated in the proposed documentary function.
The use of the \tsc language provides us flexibility in where type annotations are placed: either in all annotatable locations, or in none at all.

Two main code comprehension tasks are used in the study: code summarization and bug localization.
These two tasks are places where, subjectively, developers may state that type annotations are particularly helpful.
In the case of code summarization, type annotations can help to clarify poorly written code by providing type information for parameters with overly generic names or clarifying the scope of names.
Similarly, in bug localization, type annotations may provide information that can reveal obvious mistakes or indicate where else to look in the code (\textit{i.e.,} the definition of the type).
For each of these tasks, participants saw four code samples, which were found from GitHub,\footnote{\url{https://github.com/}} RosettaCode,\footnote{\url{https://rosettacode.org/wiki/Rosetta_Code}} or developed specifically for this experiment. 

We recruited a total of 26 participants, the majority of whom were undergraduate students. Due to technical issues (incompleteness or lack of recording) during the study, we were able to use data for 22 participants.
From their eye tracking, problem response, and survey data, we calculated various eye tracking metrics~\cite{ETmetrics}, including dwell time, fixation count, and reference behavior, which were analyzed to address the overall goal and research questions. 
We list below the overall findings and research contributions. 

In addition to the main set of tasks, we also ask participants to complete the operation span memory task~\cite{unsworth05}, a simple measure of working memory which asks participants to recall letters in order after checking if arithmetical equations are correct. We wanted to measure working memory capacity to see if it correlates with task performance. Our major findings include:

\begin{itemize}
\item The presence of type annotations does not appear to correlate with differences in reference behavior.
\item Type annotation presence and higher working memory are correlated with a slight modification of reference behavior, but together, the change is almost nil.
\end{itemize}

\paragraph*{Research Contributions} This work makes the following major contributions:

\begin{itemize}
\item The first eye tracking study to study developers' reading behavior for \tsc.
\item The first known study to consider the effect of type annotations on reading behavior, and the first to consider type-annotation-specific behavior. 
\item Insights into how developers read \tsc in the context of summarization and bug localization tasks and the relation to working memory.
The evidence suggests that type annotations are read infrequently and not referenced as much. 
\item A complete replication package~\cite{flint23:_replic_packag_do_devel_read_type_infor} of the study materials and de-identified eye tracking data for further research and extension in the research community. 
\end{itemize}

\section{Background and Prior Work}
\label{sec:prior-work}

A number of studies have been performed on both the issue of static and dynamic typing, as well as using eye tracking to understand comprehension and debugging behavior.
Additionally, we see several studies over the \tsc language, particularly from the Mining Software Repositories community.
We discuss each of these categories in turn, noting that thus far, we have not seen developer-focused studies that use the \tsc language.

\subsection{Studies over Static and Dynamic Typing}
\label{sec:studies-over-static}

Much work has been done on the subject of static and dynamic typing \cite{gannon77,kleinschmager12:_do,feldthaus14:_check_types_javas,fischer15:_api_types_javas_ms_visual_studio}.
This work has included early human-subjects studies which specifically looked at static typechecking and its relation to errors~\cite{gannon77}, or using the lens of psychology to start developing the field of program comprehension~\cite{sheil81:_psych_study_progr}.
In particular, recent work has found that static typing may provide some benefit to developers~\cite{mayer12:_empir_study_influen_static_type,fischer15:_api_types_javas_ms_visual_studio}.
These benefits include improved detectability of bugs, at least in JavaScript~\cite{gao17:_to_type_not_type}, and sped-up development time~\cite{mayer12:_empir_study_influen_static_type}.

One of the theorized reasons static typing has been theorized to benefit developers is that static type annotations may provide in-code documentation.
This theory has been explored in multiple ways, and through multiple lenses, such as software maintainability~\cite{kleinschmager12:_do}, or through the use of API deprecation to test developers' use of type annotations as in-code documentation~\cite{robbes12:_how_api}.
That type annotations interact with documentation was especially emphasized by \textcite{endrikat14:_how_api_api}, who found that the benefits of static typing are strengthened with explicit documentation.
More recently, \textcite{lubin21:_how} used grounded theory to study how developers use types in statically-typed functional languages, where they found that developers explicitly describe using type annotations as in-code documentation.
This theory provides us with the backing to explore whether or not developers actually read type annotations. 
This information is especially important, as it may help to explain results found by \textcite{hoeflich_etal2022:_highl_illog_kirk}, who found that developers sometimes prefer unsound type declarations.

\subsection{Studies on \tsc}
\label{sec:studies-typescript}

\textcite{feldthaus14:_check_types_javas} describes a method by which \tsc interfaces to JavaScript libraries may be checked.
In this, they found a number of errors between library initialization and light-weight static analysis.
Additionally, they found some limitations in \tsc's type system.

\textcite{fischer15:_api_types_javas_ms_visual_studio} span the studies over type discipline and studies over \tsc generally.
In particular, they study how both code completion and static typing affect programmer productivity, finding that, even without code completion, static typing has an effect on programmer productivity.

Moreover, \textcite{gao17:_to_type_not_type} performed a study that examined type-related bugs and code quality in JavaScript and \tsc.
They found that neither \tsc nor Flow was particularly effective in detecting the bugs, with only 60/400 found.
However, this study focused on the use of type systems after code was written and required that the type annotations were added manually by the researchers, rather than studying already annotated code.

Additionally, a more recent study by \textcite{hoeflich_etal2022:_highl_illog_kirk} found, contrary to previous studies, that sometimes, developers prefer unsound type declarations.
This is particularly helpful, as we utilize some of the same data they studied, the DependentlyTyped repository, to provide type information to our participants. This reinforces the theory that type annotations serve as in-code semantic documentation.
Similarly, \textcite{jesse_devanbu2022:_manyt} describe another large dataset of type inference information for the \tsc language, usable for training LLM tools.
Their corpus provides evidence that human annotation is valuable, and that annotation happens most frequently in \tsc on variables and parameters, the latter being an important part of in-code documentation.

More recently, \textcite{bogner_merkel2022:_to_type_not_type} examined the connection between software quality and the use of the \tsc language (considered as a statically-typed, type-safe superset of JavaScript).
They found that \tsc code was higher quality and more understandable, but not significantly less bug-prone than JavaScript itself; this increased understandability may be a result of the theorized impact of type annotations as in-code documentation.

\subsection{Eye Tracking Studies in Program Comprehension}
\label{sec:eyetr-compr-debugg}

A majority of eye tracking studies have been conducted in Java~\cite{grabinger_etal2025:_cookb_eye_track_softw_engin, sharafi_etal2015:_system_liter_review_usage_eye_softw_engin} with a few done in Python and C++~\cite{turner_etal2014:_eye_study_asses_compr_c,mansoor_etal2024:_asses_effec_progr_languag_task}.
We list a few notable studies and refer the reader to~\textcite{obaidellah_etal18:_survey_usage_eye_track_comput_progr, sharafi_etal2015:_system_liter_review_usage_eye_softw_engin, grabinger_etal2025:_cookb_eye_track_softw_engin} for a listing of eye tracking studies done within the program comprehension community. 

\textcite{kevic15:_tracin} presented an early study on how developers read code when performing change tasks.
They found that developers tend to look at and use variable names following them based on the context of the task, rather than using call-graph related constructs.
This knowledge informs our questions and design, as we also look at development change tasks, though in much smaller code, and are concerned with type annotations as well.

\textcite{rodeghero15:_empir_study_patter_eye_movem_summar_tasks} similarly performed an early study on how programmers read code for summarization.
They perform a qualitative and quantitative analysis on developer reading patterns, finding that developers read code in a left-to-right order but that there is more skimming and disorderly scanning.
We attempt to understand if there is an effect on this skimming and scanning behavior due to the presence of type annotations.

\textcite{fakhoury18} studied developer understanding of code, in particular in the presence of linguistic anti-patterns.
This was done through both the use of fNIRS and eye tracking, correlating cognitive load with identifiers and lexical features.
This study is limited only to Java code and has a different focus, yet its use of eye tracking is particularly instructive, and it also focuses on code comprehension.
Our focus is instead on other factors in code comprehension.

Both \textcite{abid-icse19} and \textcite{wallace_etal2025:_progr_visual_atten_durin_contex} conduct context-aware eye tracking studies using the iTrace framework~\cite{guarnera18} to show how developers summarize methods when provided the entire system and not just the method to be summarized. 
\textcite{karas_etal2024:_tale_two_compr} also conducted an eye tracking study on code summarization, but had participants do two uniquely different tasks --- read pre-written summaries and write their own summaries. They find that writing a summary influences where programmers focus in the code. 
Several other researchers have explored to use of eye tracking in neural models. \textcite{bansal_etal2023:_towar_model_human_atten_from} used eye movements to augment a baseline neural code summarization model (based on human attention), and \textcite{zhang-fse25} used human attention to train code LLMs. 

Other researchers have conducted eye tracking studies on assessing code readability rules for Java~\cite{park_etal2024:_eye_track_study_asses_sourc}, determining the effect of task and language (C++ vs. Python) on student comprehension~\cite{mansoor_etal2024:_asses_effec_progr_languag_task}, and studying navigation strategies in a three-phase model (finding, learning, editing) during bug fixing~\cite{sharafi_etal2022:_eyes_code}.

To date, we are not aware of a study that assesses the usefulness of type annotations using an eye tracker.
This paper is a first step in the direction of determining if developers actually read type annotations, which can help to better design tooling and languages.

\section{Methodology}
\label{sec:methodology}

The study is designed by following the structure presented in 
\textcite{grabinger_etal2025:_cookb_eye_track_softw_engin} and \textcite{sharafi_etal2020:_pract_guide_conduc_eye_track}. We now present details about the participants, tasks, metrics, and study instrumentation. 

\subsection{Participants}
\label{sec:participants}

We recruited a total of 26 qualified participants (summarized in \Cref{tab:part-desc}), primarily undergraduate students, but a handful of graduate students and working professionals (not counted separately) participated as well.
Student participants were offered \$25 Amazon gift cards and recruited through posters, announcements in courses, and through the departmental newsletter.
Professional participants were recruited through departmental connections and were offered a \$50 Amazon gift card for completion of the study.
We determined qualifications at initial contact time, when a participant reached out to schedule, we verified technical eligibility by verifying their age and experience with Javascript or \tsc before scheduling a participation time.
Of the recruited participants (for which we collected at least partial data), four must be completely suppressed, as they were unable to successfully complete eye tracker calibration (an early step after initial data collection and consent, described below).
A further five participants have incomplete data due to various technical difficulties; four of these have only one unrecorded task, with the remaining participant unable to record two bug localization tasks.
This leaves a total of 22 participants with data that can be correctly analyzed for our study.

\begin{table}[htbp]
  \centering
  \caption{Summary of participant educational backgrounds and experience levels}
  \label{tab:part-desc}
\begin{tabular}{llr}
\toprule
\multirow[c]{3}{*}{\textbf{enrolled}} & total & 26    \\
\textbf{} & suppressed & 4    \\
\textbf{} & incomplete & 5    \\
\textbf{n} &  & 22    \\
\midrule
\multirow[c]{2}{*}{\textbf{Years Programming}} & mean & 4.57 \\
\textbf{} & std & 2.31 \\
\midrule
\multirow[c]{7}{*}{\textbf{Educational Attainment}} & High School & 4    \\
\textbf{} & Some College & 10    \\
\textbf{} & Associate's & 0    \\
\textbf{} & Undergraduate & 7    \\
\textbf{} & Master's & 1    \\
\textbf{} & Doctoral & 0    \\
\textbf{} & Professional & 0    \\
\midrule
\multirow[c]{5}{*}{\textbf{JavaScript Experience}} & Very Inexperienced & 1    \\
\textbf{} & Inexperienced & 1    \\
\textbf{} & Somewhat Experienced & 15    \\
\textbf{} & Experienced & 4    \\
\textbf{} & Very Experienced & 1    \\
\midrule
\multirow[c]{5}{*}{\textbf{Typescript Experience}} & Very Inexperienced & 6    \\
\textbf{} & Inexperienced & 4    \\
\textbf{} & Somewhat Experienced & 5    \\
\textbf{} & Experienced & 6    \\
\textbf{} & Very Experienced & 1    \\
\bottomrule
\end{tabular}
\end{table}

\subsection{Tasks}
\label{sec:tasks}

Participants complete two types of tasks, code summarization and bug localization.
For each of these, they complete several subtasks (stimuli), prepared in both type-annotated and non-type-annotated form.
We detail how these were prepared below.

\subsubsection{Task Selection}
\label{sec:task-selection}

To locate task code (stimuli) for this study, we started first by searching the GitHub Gists service for self-contained, realistic code (\eg, scripts, simple libraries, or demos).
This provided a large selection of possible code, which we sorted through for size, use of external libraries and use of other features.
This allowed us to settle on our comprehension tasks, which come from a broad range of use cases, and use a broad range of features and development styles.
We located code for bug localization tasks similarly, though we searched through pull requests instead.
Finally, we used seeded bugs in two cases to ensure a broad representation of types and causes of bugs.

\subsubsection{Code Comprehension}
\label{sec:code-comprehension}

The code comprehension tasks are taken from various public GitHub repositories or Gists,\footnote{\url{https://gist.github.com}} and were re-formatted to have the same indentation and style.\footnote{This was done using the \texttt{prettier} package (\url|https://prettier.io|) with the options \texttt{prettier --no-semi -w .}.}
We then ensured that each variable declaration had a valid and usable type annotation to give the type-annotated version of the code, finally removing type annotations to provide our non-type-annotated version.
Tasks include batching item retrieval from DynamoDB (\texttt{batch\_get.ts}), conversion of ESLint results to Tool Results (\texttt{convertResults.ts}), an implementation of file operations (\texttt{cp.ts}), and an implementation of a 2-dimensional vector (\texttt{Vector2.ts}), which are further described in \cref{tab:desc-comprehension}.
The first two tasks were selected to show real-world code, albeit short; the latter show differing development styles.
In particular, the implementation of file operations uses the \enquote{lodash}\footnote{\url{https://github.com/lodash/lodash}} library which uses a highly functional style, while the 2-dimensional vector uses very object-oriented, class-forward style.
Information about the source of the code comprehension tasks is available in our replication package \cite{flint23:_replic_packag_do_devel_read_type_infor}.

\begin{table*}
  \centering
  \caption{Description and statistics of code summarization tasks.
  (\#L: lines, \#S: Statements, \#D: Declarations, \#A: Type annotations initially present, \+A: Added type annotations, Async? Asynchronous code present)}\label{tab:desc-comprehension}
\begin{tabularx}{\linewidth}{lc@{ / }c@{ / }c@{ / }c@{ / }ccX}
  \toprule
  \textbf{Name}              & \textbf{\#L} & \textbf{\#S} & \textbf{\#D} & \textbf{\#A} & \textbf{+A} & \textbf{Async?} & \textbf{Description}                                                                                                                                                                                                                                     \\
  \midrule
  \texttt{Vector2.ts}        & 56           & 20           & 11           & 15           & 3           & No              & Represents a 2D vector with x and y components and provides methods for basic vector operations such as addition, multiplication, dot product, normalization, and inversion.  (GitHub Gist) \\
  \texttt{batch\_get.ts}     & 80           & 21           & 10           & 11           & 4           & Yes             &  Provides utility functions for batch-retrieving items from an AWS DynamoDB table. It splits large key lists into chunks of 100 (the DynamoDB limit), fetches them in parallel, and optionally filters out missing or undefined results.  (GitHub) \\
  \texttt{convertResults.ts} & 39           & 15           & 10           & 3            & 9           & No              & Converts ESLint linting results into ToolResult objects by transforming errors and warnings into either FileError or Issue instances, filtering out blacklisted rules, and optionally computing fix suggestions.  (GitHub) \\
  \texttt{cp.ts}             & 95           & 38           & 20           & 28           & 4           & No              & Defines a set of asynchronous file operations that recursively copy files and directories. It provides utility functions (`cp`, `cpIf`, `copyDir`, `copyFile`, etc.) to copy files or entire directory trees while using observables to handle asynchronous filesystem tasks like reading, writing, and directory traversal.  (GitHub Gist) \\
  \bottomrule
\end{tabularx}
\end{table*}

\begin{table*}
  \centering
  \caption{Description of solutions for bug localization tasks.
  (\#L: lines, \#S: Statements, \#D: Declarations, \#A: Type annotations initially present, \+A: Added type annotations, Async? Asynchronous code present)}\label{tab:desc-localization}
\begin{tabularx}{\linewidth}{lc@{ / }c@{ / }c@{ / }c@{ / }cccX}
  \toprule
   \textbf{Name}               & \textbf{\#L} & \textbf{\#S} & \textbf{\#D} & \textbf{\#A} & \textbf{+A} & \textbf{Async?} & \textbf{Line} & \textbf{Solution Description}                                                                                                                                                                     \\
  \midrule
  \texttt{avl-tree.ts}         & 212          & 91           & 24           & 29           & 0           & No              & 8             & Strict comparison prevents insertion, because the right and left fields of the node are undefined.  Initializing them in the constructor (after line 8), should fix this.  (Rosetta Code)                       \\
  \texttt{hash-table.ts}       & 51           & 19           & 17           & 19           & 1           & No              & 33            & On deletion, \texttt{bucket[index]} is set to null, but the existence checker assumes that it should be undefined.  Replacing null with undefined will fix.  (Custom Written)                                     \\
  \texttt{vscode-wordcount.ts} & 88           & 27           & 18           & 3            & 6           & No              & 51            & When splitting an empty string, there will still be one element in the resulting list.  A check for this condition is necessary.  (GitHub)                                                                \\
  \texttt{juliaDebug.ts}       & 594          & 134          & 49           & 55           & 35          & Yes             & 278           & The \texttt{args.noDebug} and \texttt{args.stopOnEntry} variables are nullable, so where they are used, they should be \texttt{false} otherwise, by appending \texttt{?? false} to the use sites. (GitHub) \\
  \bottomrule
\end{tabularx}
\end{table*}

\subsubsection{Bug Localization}
\label{sec:bug-localization}

We selected several code from several open-source \tsc projects, in particular, VSCode plugins.
We identified candidate files through the use of GitHub search: first for VSCode-related projects written in \tsc, then through looking for single-commit fixes in \emph{merged} PRs which have small diffs (single statement) in one code file, or diffs of the same format in relatively few code files (\(\leq 3\), tests not included), our final task selection is described in \cref{tab:desc-localization}.
Additionally, to reduce the impact of bug type, we chose two which mave more explicitly type-related bugs (\texttt{juliaDebug.ts} and \texttt{hash-table.ts}), and two which have semantically oriented bugs (\texttt{vscode-wordcount.ts} and \texttt{avl-tree.ts}).
For each buggy file, we performed several steps to normalize the code.
First, we removed any unused imports, then, using the \enquote{package.json} file, we located type definitions for any imported module.
Then, we ensured that all type annotations are present, using type hints from the VSCode editor.
Next, we performed code reformatting using \texttt{prettier} as described above.
We then prepared an unannotated variant manually.

\subsection{Areas of Interest}
\label{sec:areas-interest}

To answer our research questions, we consider only line-level areas of interest (AOIs), specifically lines that contain declarations or type annotations.
In this study, we use line-level AOIs versus token-level AOIs to ensure that we capture all fixations on the line containing the declarations and annotations.  Oftentimes, a line may contain the annotation looked at through our peripheral vision, which might not be captured if token-level AOIs are used. 
An example of our training task is shown in \cref{lst:aois}.

\begin{lstlisting}[language=TypeScript,float,caption={[Training code with areas of interest highlighted.]Training code with areas of interest shown in \aoibox{orange}, and type annotations shown in \annotbox{blue}.\vspace{-2em}},label={lst:aois}]
(*@\aoibox{\textbf{let} i: \annotbox{\textbf{number}}}@*)
(*@\aoibox{\textbf{let} output: \annotbox{\textbf{string}}}@*)

for (i = 1; i <= 100; i += 1) {
  output = ""
  (*@\aoibox{\textbf{let} divBy3: \annotbox{\textbf{boolean}} = !(i \% 3)}@*)
  (*@\aoibox{\textbf{let} divBy5: \annotbox{\textbf{boolean}} = !(i \% 5)}@*)

  if (divBy3) output += "Fizz"
  if (divBy5) output += "Buzz"

  // Converts to a string because output is defined as such
  if (!(divBy3 || divBy5)) output = i.toString()

  console.log(rot13(output))
}

(*@\aoibox{\textbf{function} rot13(str: \annotbox{\textbf{string}}): \annotbox{\textbf{string}}: \{}@*)
  return str
    .split("")
    (*@\aoibox{.map((char: \annotbox{\textbf{string}}) =>}@*)
      String.fromCharCode(
        char.charCodeAt(0) + (char.toLowerCase() < "n" ? 13 : -13),
      ),
    )
    .join("")
}
\end{lstlisting}

\subsection{Variables}
\label{sec:variables}

We consider three categories of variables, independent variables, control variables (measures which we can statistically control for), and dependent variables; these are shown in \cref{tab:study-vars}.
Each subject and dependent variable (and its measurement) is described below.

\begin{table}[htbp]
  \centering
  \caption{Study Variables and Roles.
  These are described in \Cref{sec:variables}.}
  \label{tab:study-vars}
\begin{tabular}{lll}
  \toprule
  \textbf{Independent} & \textbf{Control}   & \textbf{Dependent}         \\
  \textbf{Variables}   & \textbf{Variables} & \textbf{Variables}         \\
  \midrule
  Annotation Status    & Experience         & Correctness                \\
  Code Complexity      & Education          & Time on Task               \\
  Stimulus (Code Shown)                     & Working Memory     & \textbf{Eye Metrics:} \\
                       &                    & Time in AOI                \\
                       &                    & \% Time in AOI             \\
                       &                    & Fixations on AOIs          \\
                       &                    & \% Fixations on AOIs       \\
                       &                    & Regressions to AOI         \\
  \bottomrule
\end{tabular}
\end{table}

\begin{description}
\item[Experience] 
  We follow the pattern of \textcite{siegmund13:_measur} for self-estimation of programming experience in the JavaScript and \tsc languages.
  In particular, we ask for the level of experience on a five-point Likert-type scale (from \enquote{Very Inexperienced} to \enquote{Very Experienced}), in general, and as compared to peers.
  Additionally, we ask how long developers have been programming and how long they have been programming in the last six years.
\item[Education]
  We collect information about each participant's educational background.
  In particular, we ask for their highest level of educational attainment, ranging from high school to graduate and professional degrees.
\item[Correctness] 
  After each task, participants answer one or two questions about the code they were shown.
  For basic summarization tasks, participants are asked to describe the code.
  We consider correctness as binary, whether the description was accurate.
  This was done because participant responses were short and we wanted to avoid the issue of determining credit for partial correctness.

  For bug localization, they are asked to state what lines need fixed and to describe the cause of the bug.
  The description of the bug cause is rated in the same way as the description of the code for comprehension tasks.
  Correctness of fixing lines is binary; either the participant selected the correct line or did not.
\item[Time on Task]
  We measure time on task as the time from the first gaze on the target code to the time tracking stops.
\item[Working Memory (Partial Credit Load)]
  We measure working memory using Partial Credit Load, a derived metric which considers partial recall weighted by the load at time of recall~\cite{conway_etal2005:_workin_memor_span_tasks}.
\item[Time in AOI (Cumulative, Percentage)]
  We consider time in AOI as the cumulative time spent in a particular AOI or AOI category, as the percentage of time spent, or the mean time spent in a particular category.
\item[Fixations in AOI (Raw, Percentage)]
  Number or percentage of fixations on a particular AOI or AOI category.
\item[Regressions to AOI]
  Number of fixations which are the result of re-reads or re-visits (upward vertical or leftward horizontal movements) to an AOI.
\end{description}

\subsection{Eye Tracking Apparatus and Environment}
\label{sec:eyetr-instr}

We use the iTrace infrastructure \cite{guarnera18} to collect eye movement data.
This is built in two parts, the \textit{core} and the \textit{plugin}.
The core is responsible for overall management of eye tracking, including calibration, connection to the eye tracker, and study/task metadata.
The plugin works with an IDE or development environment to collect information about Areas of Interest (AOIs) at different levels of granularity.
This information is collected as file, line, and column numbers, which are post-processed using the iTrace toolkit \cite{developers21:_toolk}, which detects fixations and provides access to information about them through an SQLite database.

Tasks were presented on a Windows 10 computer with a screen resolution of \qtyproduct{1920 x 1080}{\px} running version 1.60.0 of the Atom text editor,\footnote{\url{https://atom.io}} using the Atom iTrace plugin.\footnote{\url{https://github.com/iTrace-Dev/iTrace-Atom}}
The Atom environment was configured using the Atomic Management Plugin\footnote{\url{https://atom.io/packages/atomic-management}} to use the \texttt{atom-light-syntax} and \texttt{atom-light-ui} themes with \qty{24}{\px} fonts to reduce effects of eye strain on tracking results.
Gaze was tracked using the Tobii Spectrum eye tracker running at \qty{120}{\hertz}.
To calculate fixation information, we used the IDT filter with a duration of \qty{10}{ms} and a dispersion of \qty{125}{\px}~\cite{AnderssonOneAlgoToRuleAll}.
Fixations were mapped using a custom script (found in \textcite{flint23:_replic_packag_do_devel_read_type_infor}), which used abstract syntax tree information from the \tsc compiler to determine fixated-upon tokens.

\subsection{Analysis}
\label{sec:analysis}

First, to explore \ref*{rq:change-reading-behavior}, we use a between-groups ANOVA (omnibus \(F\)) to determine if there exists a difference in any analyzed factor.
From the \(F\)-test, we follow up with Tukey's HSD, a self-adjusting pairwise test of differences.
Cases with missing data were not removed, leaving imperfect data, but this fits given the underlying mechanics of the ANOVA.
We test for differences in the count of fixations on AOI, percent time spent in AOI, and percent fixations on AOI, for only regression fixations, and for all fixations.
For each of these metrics, it is not possible to detect outliers due to the relatively small sample size and time limitations given to participants.
We consider the effect of stimulus (that is which piece of code is shown), of type annotation presence, and the interaction as well.
We performed counterbalancing with respect to the combination of stimulus and type annotation presence, but to avoid learning effects, we used randomized presentation after the counterbalancing process.
As relevant, we report the results of the \(F\)-test, the results of the HSD for a particular pair, and visualizations as relevant.

Then, to examine the differences in correctness between stimulus (\ref*{rq:correctness-behavior}), we first have one author manually label each response as (in)correct.
Following the guidance of \textcite{ahmed2024}, we have four LLMs (\texttt{gpt-4.1-mini-2025-04-14}, \texttt{gpt-4.1-2025-04-14}, \texttt{gemini-2.0-flash-001}, and \texttt{gemini-2.5-flash-preview-05-} \texttt{20}, chosen to provide a mixture of training data while also keeping costs low) also label the data and compute model-model agreements.  Here, we prompted the LLMs with the code/bug description and the human response and asked it if they are similar or different.  For both tasks, the model-model agreement levels were above 0.5 (Krippendorff's alphas of \statistic{\alpha}{0.544} and \statistic{\alpha}{0.612}, respectively), indicating that an LLM can replace one human rater~\cite{ahmed2024}.

When checking the agreement of the best performing LLM (\texttt{gpt-4.1-2025-04-14}) with the human rater, there was high agreement (\statistic{\alpha}{0.703}) for the localization correctness task, while for the comprehension correctness task, there was very low agreement (\statistic{\alpha}{0.184}).  The LLMs seemed to be overly pessimistic, rating a majority of responses as incorrect.  For this task, we added a second human rater.  Both human raters have over 8 years of experience in software engineering.  Here, each rater gave a binary yes/no decision on if the participant responses were similar (or not) to the given description.  The agreement before discussion was \statistic{\alpha}{0.421}.  After a discussion round, where any disagreements were discussed and each rater had the opportunity to change their rating (but did not have to), the agreement was much higher at \statistic{\alpha}{0.902}.

We then use Cochran's \(Q\), an omnibus test for repeated-measures binary variables.
Then, we follow up with a pairwise Bonferroni-adjusted McNemar's \(\chi^{2}\) to determine differences between each stimulus.
Finally, we use Fisher's Exact Test to determine if the type annotation state influenced the number of correct responses.

Next, we examine the relationship between working memory and reference to type annotation and declaring lines (\ref*{rq:improve-performance-low-wm}).
We do this using standard ordinary least squares regression, combining a working memory measure (the partial credit load) and type annotation status to model percent time in AOI and percent fixations on AOI.
We report details of each model and show a basic visualization of the more interesting models.

Finally, we consider developer opinions on type annotations (\ref*{rq:prefer}).
For this, we report response summaries on several questions and use Fisher's Exact Test with Bonferroni correction to determine if associations between questions exist.

\subsection{Procedure}
\label{sec:procedure}

When a participant arrived at their scheduled time, we first briefed them on the procedures of the study, including all portions related to risks, data collection, and informed consent. We first collected documentation of informed consent.
Participants then completed a brief pre-study questionnaire about their educational and programming experience, followed by completion of the operation span working memory task.
We then calibrated the eye tracker, at this point, if a participant was unable to complete calibration after three tries, they were dismissed.
Next we, trained participants on the study procedures using the FizzBuzz task, and continued to record the code summarization task.
After every second stimulus, we would re-calibrate the eye tracker and then continue to make sure we minimized drift.
At this point, we performed a short, structured interview to understand the participant's overall strategy, and participants were given a moment to stand and take a brief break.
We would then repeat the same process for the bug localization tasks.
Finally, we collect self-report information about programming experience (in general, compared to peers, and for the JavaScript and \tsc languages, see also \cite{siegmund13:_measur}), as well as participant opinions and experiences of type annotations, as part of a post-study questionnaire.
We then collected any relevant information about payment or extra credit and thanked the participant for their time.

\section{Research Questions}
\label{sec:research-questions}

The study seeks to answer the following research questions. 

\begin{rqs}
\item \label{rq:change-reading-behavior} \textbf{Does type annotation presence change reading behavior?}
  In particular, are there more or fewer regressions to lines containing declarations or type annotations?
  Are there differences in how many or how long participants fixate on lines containing declarations or annotations?
  If so, this should help to show the impact of type annotations as in-code documentation.
\item \label{rq:correctness-behavior} \textbf{Is there a relationship between task correctness and reading behavior with respect to type annotation presence?}
  That is, do correct responses see more or fewer regressions to lines containing declarations or type annotations when type annotations are present.
\item \label{rq:improve-performance-low-wm} \textbf{Is there a correlation between reference behavior (regressions to declaring/annotated lines) and working memory scores?}
  In particular, this would support previous theories that type annotations are a memory aid to developers.
\item \label{rq:prefer}\textbf{Do developers prefer type annotations or find them to have utility?} We provide answers to developer preferences via a post-study survey on type preference. We do this to determine preferences compared to actual behavior (shown in the first three RQs).
\end{rqs}
In this study, we consider two different categories of task, code summarization and bug localization, which are described in greater detail in \cref{sec:tasks}. Except for \ref*{rq:prefer}, which is answered in the post-study survey, we examine research questions 1 through 3 for each of the task categories.

\section{Results}
\label{sec:results}

We present the results for each of the four research questions and highlight the findings for each. 

\subsection{\ref*{rq:change-reading-behavior}: Reading Behavior}
\label{sec:rq-change-reading-behavior}

\begin{figure*}[htbp]
  \centering
  \begin{subfigure}[t]{0.45\linewidth}
    \centering
    \includegraphics[width=\linewidth]{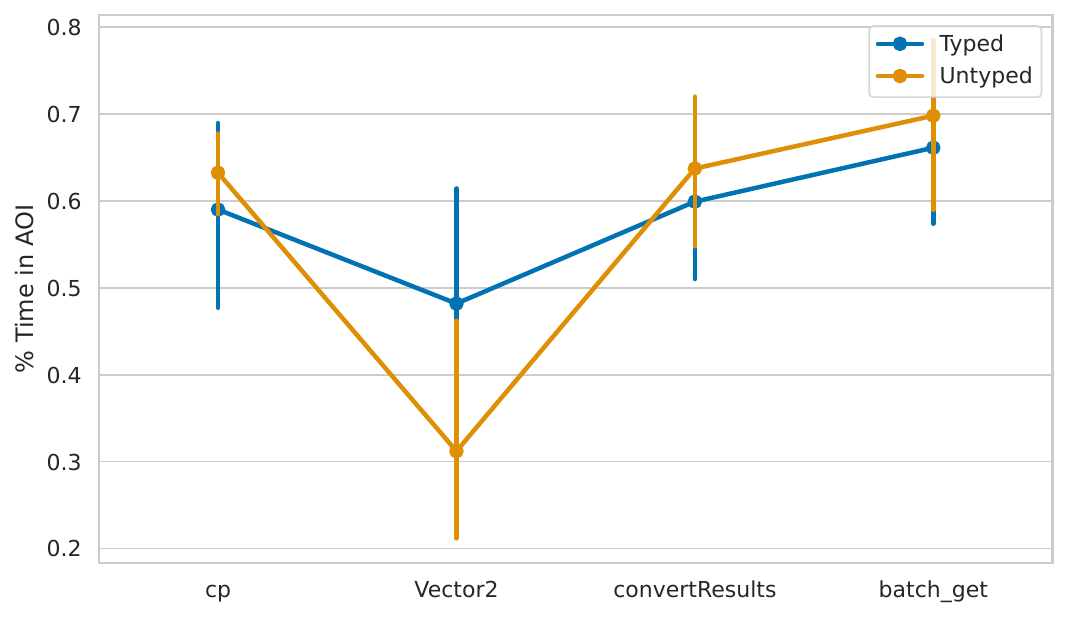}
    \caption{Code Comprehension}\label{fig:linearity-comprehension}
  \end{subfigure}
  \begin{subfigure}[t]{0.45\linewidth}
    \centering
    \includegraphics[width=\linewidth]{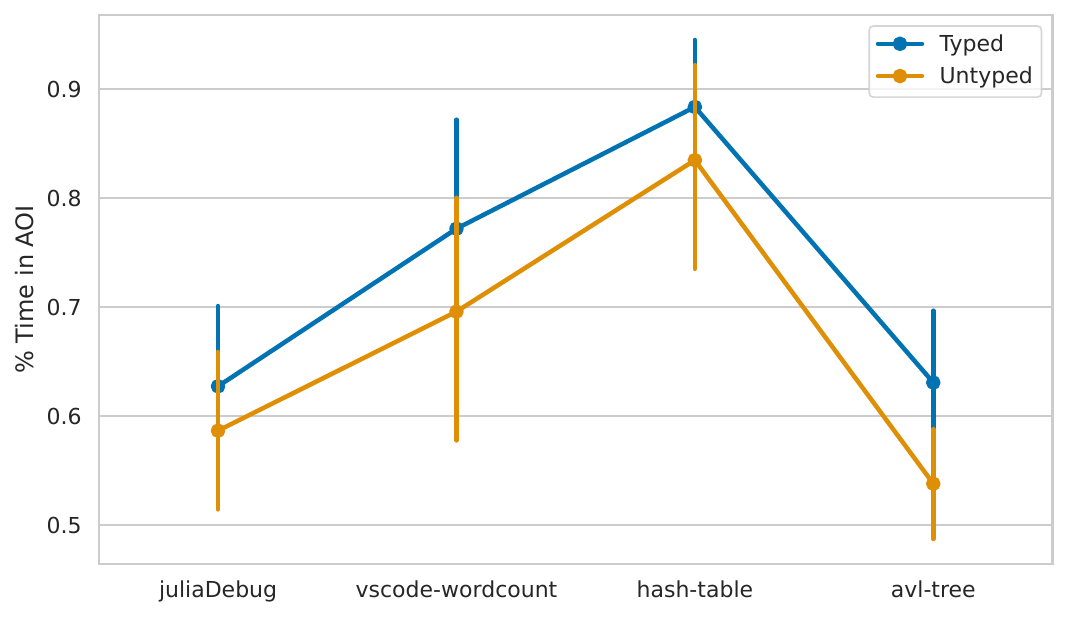}
    \caption{Bug Localization}\label{fig:linearity-localization}
  \end{subfigure}

  \caption[Differences between time spent on AOIs.]{Differences between time spent on AOIs. The \(y\)-axis shows the percent of time participants spent in AOIs.}
  \label{fig:diff-time-aoi}
\end{figure*}

First, we consider how the presence of type annotations impacts developer reading behavior, in particular, related to reference behavior to lines containing type annotations or declarations.
Using fixation-derived metrics of time in AOI and number of fixations in AOI, we first look at how much developers refer to type annotations or declarations; a graph of time spent on AOIs is shown in \cref{fig:diff-time-aoi}.
For both overall tasks, we see that any difference is primarily from the specific code shown, though in the code comprehension task (\cref{fig:linearity-comprehension}), type annotations may have a slightly stronger impact (\pvalue{0.07139237}, \Fmeasure{1}{80}[3.338817]).
Not only do the statistical results tell us that developers did not spend significant time reading annotatable locations, but that their presence does not impact those locations being read.

\begin{finding}{}{fd:not-read}
  Developers do not appear to read type annotations or type-annotatable areas more often when type annotations are present.
  Moreover, they do not appear to read these areas frequently to begin with.
\end{finding}

Additionally, regressions to AOIs (an indicator of possible reference behavior) were infrequent.
Differences in the number of regressions to AOIs were primarily driven by stimulus, rather than type annotation presence (for comprehension, \Fmeasure{3}{55}[3.262261], \pvalue{0.028151}; for localization \Fmeasure{3}{75}[5.02667], \pvalue{0.003145}).
That is, participants did not appear to refer back to type annotations, and we discuss possible reasons for this later.

\begin{finding}{}{fd:not-referred}
  Not only are type annotations read infrequently, but in our context, they were not regressed to.
  That is, developers did not appear to use type annotations as in-code documentation or reference material.
\end{finding}

\subsection{\ref*{rq:correctness-behavior}: Correctness and Reading Behavior}
\label{sec:correctness-behavior}

Next, we consider the effect of annotation presence on correctness.
For both tasks, we find that the presence of type annotations has no impact on correctness (for comprehension, \pvalue{1.0}; for localization, \pvalue{0.6176604027144509}).
This is unsurprising, given prior research on the impact of type annotations on downstream task correctness, which has found an inconsistent effect~\cite{prechelt98:_contr_exper_asses_benef_proced,okon16:_can}.

On the other hand, we do see for both tasks an impact of the stimulus.
In the case of comprehension, the impact is significant (\statistic{Q(3)}{10.959183673469388}, \pvalue{0.011948644758529959}); however, we do not see an adjusted difference between any given pair of stimuli.
Similarly, when we consider developers naming the fixing line for bug localization, we see a global difference (\statistic{Q(3)}{25.58823529411765}, \pvalue{1.163105756259275e-05}) and a difference between several pairs: avl-tree/hash-table, \pvalue{0.002930} and hash-table/juliaDebug \pvalue{0.001465}.
We see a similar pattern for providing a description of the cause of the error: a global difference exists (\statistic{Q(3)}{24.134328358208954}, \pvalue{2.341774174223858e-05}), and we see several differences between pairs: avl-tree/hash-table (\pvalue{0.001465}), avl-tree/vscode-wordcount (\pvalue{0.010986}), and hash-table/juliaDebug (\pvalue{0.044312}).

\begin{table}[htbp]
  \centering
  \caption{Summary of response correctness for all participants.}
  \label{tab:response-correctness}
  \begin{subtable}[t]{\linewidth}
    \centering
    \caption{Participant correctness for comprehension tasks. 
    \(C / T\) represents \(C\) correct responses across \(T\) total participants seeing the stimulus in that form.}
\begin{tabular}{lr@{\ /\ }lr@{\ /\ }l}
\toprule
 & \multicolumn{2}{r}{\textbf{Annotated}} & \multicolumn{2}{r}{\textbf{Unannotated}} \\
\textbf{Vector2} & 10 & 10 & 9 & 12 \\
\textbf{batch\_get} & 6 & 12 & 5 & 10 \\
\textbf{convertResults} & 6 & 10 & 8 & 12 \\
\textbf{cp} & 6 & 12 & 6 & 10 \\
\bottomrule
\end{tabular}
  \end{subtable}
  \begin{subtable}[t]{\linewidth}
    \centering
    \caption{Participant correctness for localization tasks. 
    \(L / D / T\) represents \(L\) participants who selected the correct line, \(D\) who provided an accurate description of the cause of the bug, and \(T\) who saw that stimulus.}
\begin{tabular}{lr@{\ /\ }c@{\ / \ }lr@{\ /\ }c@{\ / \ }l}
\toprule
 & \multicolumn{3}{r}{\textbf{Annotated}} & \multicolumn{3}{r}{\textbf{Unannotated}} \\
\textbf{avl-tree} & 0 & 0 & 5 & 1 & 1 & 17 \\
\textbf{hash-table} & 8 & 8 & 12 & 5 & 6 & 10 \\
\textbf{juliaDebug} & 0 & 3 & 17 & 0 & 0 & 5 \\
\textbf{vscode-wordcount} & 1 & 5 & 10 & 6 & 8 & 12 \\
\bottomrule
\end{tabular}
  \end{subtable}
\end{table}

\begin{finding}{}{fd:task-correctness}
  Specific stimulus more than type annotation presence impacts developer correctness.
  This is unsurprising when compared to previous work, where the effect of type annotations is somewhat unclear.
\end{finding}

\subsection{\ref*{rq:improve-performance-low-wm}: Working Memory and Reference Behavior}
\label{sec:wm-perf}

Next, we consider if there exists any relationship between working memory capacity and developers' reference behavior.
To do this, we use the percent of time spent in AOIs modeled by whether the stimulus was typed, the partial credit load, and their interaction.
We computed a separate model for both comprehension and localization, with the percent time spent in AOI considered per participant/stimulus.
The results of ordinary least squares regression for these models are shown in \cref{tab:ols-results}.

When we consider the results for comprehension (\cref{tab:ols-comprehension}), we see several interesting things.
First, we see that type annotations are associated with a decrease in the relative time spent in AOIs (\statistic[4]{b}{-0.0078}), but so is higher partial credit load, or higher working memory (\statistic[4]{b}{0.180}).
However, the combination of type presence and higher partial credit load increases the amount of time spent in AOIs, emphasizing how weak the relationships are.

The results for bug localization are slightly different (\Cref{tab:ols-localization}.
Here, we see that type annotations are associated with an increase in reference behavior (\statistic[4]{b}{0.280}), as is a higher partial credit load (\statistic[4]{b}{0.180}), though the interaction of the two reduces the amount of reference behavior.
This pattern is the opposite of comprehension, but the end result is the same: working memory does not seem to have an impact on developer reference behavior.

\begin{table*}[htbp]
  \centering
  \caption{Ordinary Least Squares Regression Results for \ref*{rq:improve-performance-low-wm}, using \(\text{time} \sim \text{typed} + \text{pcl} + \text{typed}\times\text{pcl}\).}
  \label{tab:ols-results}
  \begin{subtable}{1.0\linewidth}
    \centering
    \caption{For Comprehension}
    \label{tab:ols-comprehension}
\begin{tabular}{lcccccc}
\toprule
                                                    & \textbf{coef} & \textbf{std err} & \textbf{t} & \textbf{P$> |$t$|$} & \textbf{[0.025} & \textbf{0.975]}  \\
\midrule
\textbf{Intercept}                                  &       0.2147  &        0.071     &     3.041  &         0.002        &        0.076    &        0.353     \\
\textbf{C(is\_typed)[T.True]}                       &      -0.0778  &        0.103     &    -0.759  &         0.448        &       -0.279    &        0.124     \\
\textbf{partial\_credit\_load}                      &      -0.1048  &        0.082     &    -1.272  &         0.204        &       -0.266    &        0.057     \\
\textbf{partial\_credit\_load:C(is\_typed)[T.True]} &       0.1129  &        0.120     &     0.941  &         0.347        &       -0.123    &        0.349     \\
\bottomrule
\end{tabular}
  \end{subtable}
  \begin{subtable}{1.0\linewidth}
    \centering
    \caption{For Localization}
    \label{tab:ols-localization}
\begin{tabular}{lcccccc}
\toprule
                                                    & \textbf{coef} & \textbf{std err} & \textbf{t} & \textbf{P$> |$t$|$} & \textbf{[0.025} & \textbf{0.975]}  \\
\midrule
\textbf{Intercept}                                  &       0.0733  &        0.047     &     1.565  &         0.118        &       -0.019    &        0.165     \\
\textbf{C(is\_typed)[T.True]}                       &       0.0280  &        0.067     &     0.419  &         0.675        &       -0.103    &        0.159     \\
\textbf{partial\_credit\_load}                      &       0.0180  &        0.055     &     0.327  &         0.744        &       -0.090    &        0.126     \\
\textbf{partial\_credit\_load:C(is\_typed)[T.True]} &      -0.0371  &        0.078     &    -0.473  &         0.636        &       -0.191    &        0.117     \\
\bottomrule
\end{tabular}
  \end{subtable}
\end{table*}

\begin{finding}{}{fd:memory-reference}
  When working memory and type annotation presence is taken into account, although individually they impact reference behavior, together the effect is nearly nil.
\end{finding}

\subsection{\ref*{rq:prefer}: Developer Preferences for Annotations}
\label{sec:dev-preference}

\begin{figure*}[htbp]
  \centering
  \includegraphics[width=0.72\linewidth]{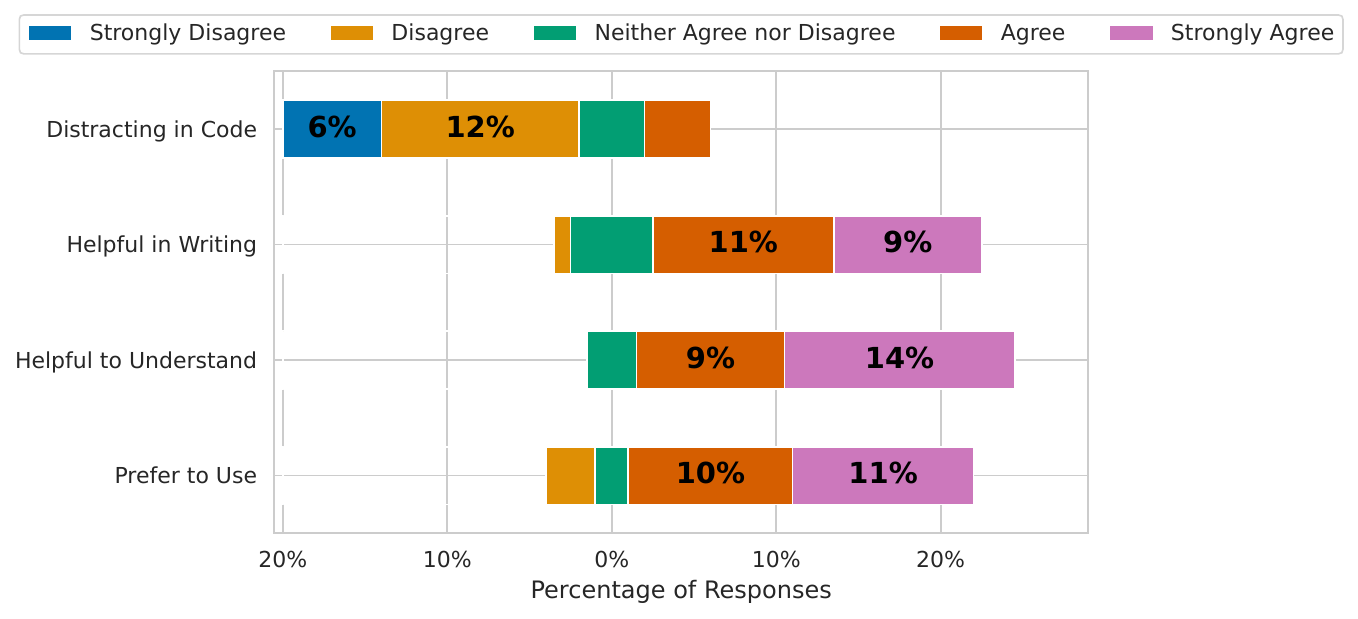}
  \caption{Participants' responses to questions about views on type annotations. 
    Questions include whether type annotations are distracting, whether they are helpful in writing code, helpful in understanding code, and whether participants prefer to use type annotations in their own code. 
    Notably, participants do not appear to find type annotations distracting, and generally find them helpful.}
  \label{fig:views-on-annots}
\end{figure*}

Finally, we show developers' self-reported preferences around type annotations.
During the post-study survey, we asked participants questions about preferences related to type annotations.
In particular, we asked about whether they prefer to use type annotations in their code, whether they find type annotations to be helpful in writing code, and in understanding code.
We additionally asked if they find type annotations to be a distraction.
We show a summary of participant responses in \cref{fig:views-on-annots}.

We find that participants generally prefer to use type annotations (21), find type annotations helpful in understanding code (23), and in writing code (18).
This follows from previous research, where developers described their type annotation practices~\cite{lubin21:_how}.
Relatedly, we also found that most participants did not find type annotations to be a distraction (18).

\begin{figure}[htbp]
  \centering
  \includegraphics[width=\linewidth]{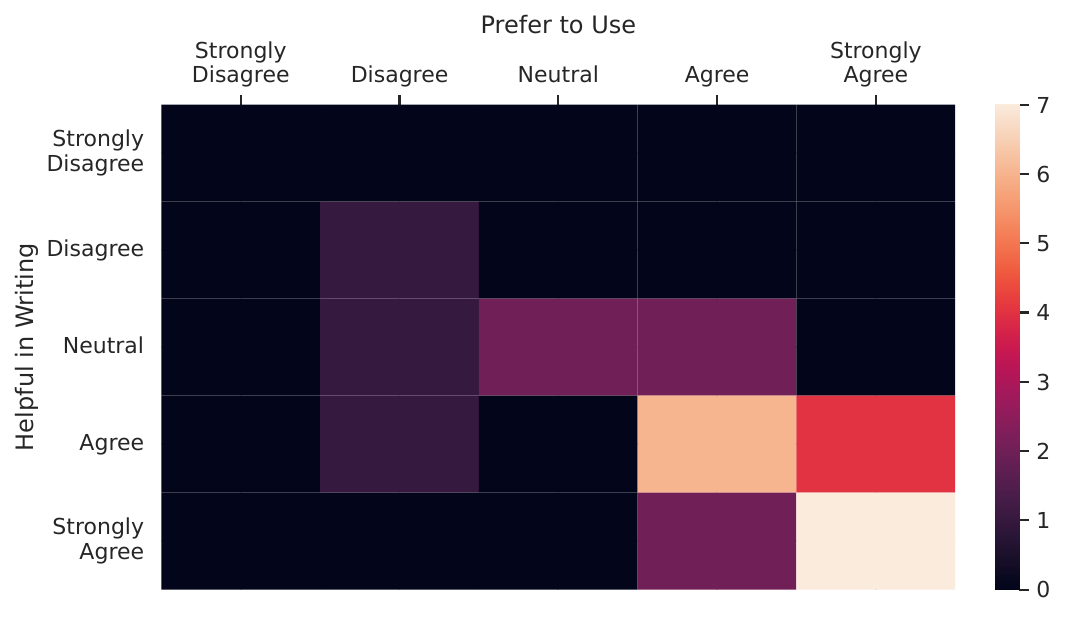}
  \caption{Heatmap showing the relationship between developers preferring type annotations and finding them helpful in writing code. 
    Participants who reported they prefer to use type annotations also reported they found them helpful in writing (\pvalue{0.0426}).}
  \label{fig:prefer-helpful-in-writing}
  \vspace{-1em}
\end{figure}

\begin{figure}[htbp]
  \centering
  \includegraphics[width=\linewidth]{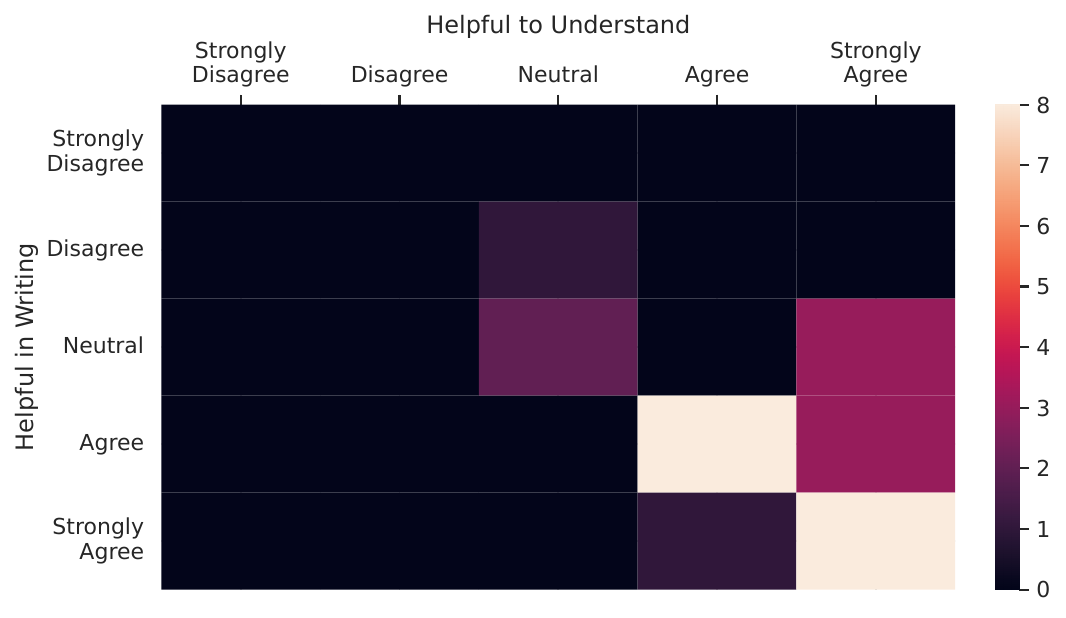}
  \caption{Heatmap showing the relationship between developers finding annotations are helpful in understanding and finding them helpful in writing code.
    Participants who reported they find type annotations helpful to understand code found them helpful in writing (\pvalue{0.0036}).}
  \label{fig:helpful-understanding-helpful-writing}
  \vspace{-1em}
\end{figure}

From these results, we performed a basic correlational analysis using Fisher's Exact Test followed by Bonferroni correction.
We found a significant association between developer preference for annotations and participants finding them helpful in writing \pvalue{0.0426} (shown in \cref{fig:prefer-helpful-in-writing}), and between finding annotations helpful in understanding and helpful in writing \pvalue{0.0036} (shown in \cref{fig:helpful-understanding-helpful-writing}).
In both cases, we see both statements tend towards mutual positive agreement, though the developers who did not prefer type annotations were not inclined to agree that they were helpful in writing code.
Additionally, although it is not significant after the correction (\pvalue{0.0816}), preference for annotations and being helpful in understanding was significant before the correction (\pvalue{0.0136}), suggesting that further research is necessary in this area. 

\begin{finding}{}{fd:preference}
  Developers view the use of type annotations positively.
  In particular, developers who preferred type annotations found them helpful in writing, or those who found them helpful in understanding found them helpful in writing.
\end{finding}

\section{Discussion and Implications}
\label{sec:discussion}

In this section, we consider possible causes of our results, both behavioral and language-level, and discuss implications for four major domains.

\paragraph{Reference Behavior and Memory}  Taken together, \Cref{fd:not-read,fd:not-referred,fd:memory-reference} help us to better understand how developers use type annotation information.
Although it is theorized to be in-code documentation~\cite{lubin21:_how}, our results contextualize this. 
Prior work has generally assumed that developers read type annotations as part of information-seeking behavior; our results show that the lines that contain type annotations are just not read that often.
Instead, while it is still likely that developers read this information, we are unable to detect it as a fixation.
Developers may have glanced at it too quickly to detect, or it may be read in their peripheral vision.

However, as the specific task being worked on drove differences in developer reference behavior, we instead would consider that more reasonably, information needs are task-oriented, and annotatable-location information often contains this, whether or not type annotations are present.
For example, given working memory scores, it may be that developers could maintain enough contextual information during their completion of these tasks.
Different contributors to task complexity (including name descriptiveness, code size, or others) are likely greater contributors to developers' use of type annotations as in-code documentation, just as they likely are contributors to differences in correctness (\Cref{fd:task-correctness}).

\paragraph{Preference} Participants in our study expressed that type annotations presented great value to them in their process (\Cref{fd:preference}), which means that even if we were unable to detect a difference in reading behavior due to them, they are still useful.
In part, this utility may be reduced in some cases by good naming practices, but type annotations present other opportunities.
Type information is in-code documentation, but tools such as IDEs can harvest this to present on-the-spot information, or it could be used to generate more complete documentation sets.
Additionally, this preference reinforces the documentary theory: that developers find type annotations to be helpful in writing and understanding code indicates that they do use them in some way.
This holds, even if we were unable to detect their use as reference information.

These overall results have implications for builders of software tools, the software development community, the education of future developers and engineers, and research in program comprehension and software engineering.

\paragraph{Tool Builders} As tooling can have an impact on developer productivity, bearing in mind developer informational needs is especially important.
While developers explicitly describe type annotation information as important, how it is used remains unclear; as such, tool builders have the opportunity to develop novel ways of providing this information or improving its accessibility.
Tools like the Language Server Protocol can be leveraged to help inform developers, which is likely the interaction that they are used to.
Further leveraging type information, and developers' interactions with this information, is therefore still incredibly important; interactions in this area include those discussed, for example, by \textcite{lubin21:_how}.

\paragraph{The Software Development Community} Similarly, how in-code type information, i.e., type annotations, are used is important among working developers.
Developers clearly see type annotations as in-code documentation, and they are likely used as such; developing style guides which help developers make appropriate decisions for including type annotations.
In particular, recognizing where this information is most needed becomes important, but so does designing good naming policies.

Additionally, language designers, as part of the software development community, may find this information useful.
Prior work has shown that type annotations are used in certain locations more often, and omitted elsewhere~\cite{flint_etal24:_how_do_devel_use_type_infer}; as new languages are developed, continuing to allow for these patterns, based not only on large-scale evidence, but on developer reference behavior, is especially important.

\paragraph{Education} Additionally, type-annotation information is important in education.
Although institutions teach programming differently, the key concept of data types is frequently taught early on.
In some languages, this is more implicit than others (\eg, Python), yet teaching students how they can use this information is important.
For example, type information can still provide information about a program's problem domain, and helping students to learn how to read code with type annotations is important.

\paragraph{Research} Finally, this research follows in a long line of work exploring the nature and impact of type systems on developers' code comprehension.
While our results strengthen the evidence that developers see type annotations as in-code documentation, how this information is gathered or utilized remains unclear.
Further work to understand this mechanism is necessary, but the benefits of static typing, absent our understanding of how type information is used, remain.

Additionally, these results help underscore our lack of understanding of developers' information needs \dash developers see type annotations as in-code documentation, but they are not used like we would expect.
Building a model of developer information needs can help us to further build better tools, and languages.
Moreover, understanding how information needs evolve over time, and as developers gain experience, is necessary; as is understanding how information needs change throughout a project's lifecycle.

\section{Threats to Validity}
\label{sec:threats-validity}

\paragraph{Construct Validity}

There are two major threats to construct validity: whether or not eye tracking measures what we purport, and the use of the operation span task. Our use of eye tracking assumes that the eye-mind hypothesis holds, that is, participants' eye movements are closely related to the cognitive processes being carried out~\cite{reichle_reingold2013:_neurop_const_eye_link}.
Although the hypothesis has utility, it is unclear to what extent eye movements correlate with the complexities of code comprehension.
Related to this is also that areas of interest were relatively small \dash around the size of a line of code; the contents of these AOIs may be readable in peripheral vision or may not be adequately captured by our eye-tracker.
This is a reasonably general threat to eye tracking, but we attempted to mitigate it by the use of a large font-size.
The operation span task is a well-validated measurement of an individual's working memory~\cite{unsworth05}, thus it is an appropriate measure.
In particular, it has high internal consistency, and the use of partial-credit scoring follows high-quality empirical evidence~\cite{conway_etal2005:_workin_memor_span_tasks}.
Additionally, the specific implementation of the task we use implements existing recommendations for conducting working memory experiments \cite{conway_etal2005:_workin_memor_span_tasks}.

\paragraph{Internal Validity}

Our study design presents reasonable internal validity, as our mixed-group design balances complete control and efficient use of subject time.
However, because of the specific nature of some of our tasks, stimulus-level differences can impact our results.
For this, we use several statistical techniques, which help to mitigate these issues and other possible confounds.
Additionally, internal validity is potentially threatened by the presence of original names in code, which may improve developers' understanding of code.
However, given the poor performance on bug localization, this was likely not the case; moreover, due to the use of external libraries, original names were necessary to preserve information about what was being referenced.

\paragraph{External Validity}

This study has two major threats to external validity: the studied population and the selected code tasks.
Because the study uses primarily students, its generalizability is potentially limited; however, much research is performed solely with students as participants, and our participants' experience in both \tsc and JavaScript was varied. Future studies with industry developers remain necessary.
We selected two types of code tasks, some of which are from existing, realistic code bases while others are more toy-like.
Although they are limited to particular application domains, limiting complete generalizability, they come from realistic code bases, minimizing this threat.

\paragraph{Conclusion Validity}

Finally, we must consider statistical conclusion validity.  Except for the ANOVAs, all statistical assumptions were met; in the case of the ANOVAs in \ref*{rq:change-reading-behavior}, there is enough robustness to non-homogeneity that the results are robust.

\section{Conclusions and Future Work}
\label{sec:conclusion}

In this paper, we explored the impact of type annotations on developers' reading behavior in the \tsc language, through the use of eye tracking on several pieces of code.
We found that type annotations do not impact developer reading behavior overall.
However, working memory capacity may impact type reference behavior, although the effect nearly evens out when type annotations are considered.
Additionally, we found that type annotations had no impact on developers' correctness in code summarization or bug localization tasks, following previous research.
We also found that developers see type annotations positively.

From these results, we see the theory that type annotations are in-code documentation retains merit (as developers describe them as being useful in understanding code), but we are unable to fully understand how they are read.
Further long-term work understanding developer information needs and mental models is needed to better contextualize these results, but they still may provide information to tool builders, the software development community, and educators.

Given our results, we believe that future work on the impact of type annotations on developers' reference behavior is necessary. We are interested in examining how developers seek out this information via qualitative approaches, including interviews, surveys, and behavioral patterns. Furthermore, we aim to explore behavior in languages with stronger or weaker type systems, as well as more complex tasks. Another aspect of future work could study how identifier naming conventions (plurality and structure)~\cite{HostECOOP-2009, Binkley2011, NewmanJSS2020} intersect with type names. Finally, we are interested in expanding to more diverse populations, including more professional and/or open-source developers.

\section*{Data Availability}

Tasks, scripts, and de-identified data are available in our replication package, \textcite{flint23:_replic_packag_do_devel_read_type_infor}.

\section*{Ethical Compliance}

This study has been approved by the \blind{University of Nebraska-Lincoln}{Blinded} Institutional Review Board as Project \#\blind{20220721897EX}{Blinded}.
Study subjects participated of their own free will, and were able to terminate their participation at any time during the course of the study.

\begin{acks}
  This work was supported in part by the U.S. National Science Foundation under grant CNS-2346327.
  We thank Alisson Ntwali and Kareem Keshk for helping find suitable code for use in the two tasks.
\end{acks}

\balance

\printbibliography

\end{document}